\documentclass[reprint]{revtex4-1}
\usepackage{epsfig,amssymb,amsmath,textcomp,subfigure,caption,verbatim,gensymb,braket}
\usepackage{epstopdf,subfig}
\usepackage{graphicx}
\usepackage{amsfonts}
\begin{document}

\rightline{DAMTP-2016-57}

\vskip 10pt

\title{A dynamical $\alpha$-cluster model of $^{16}$O}

\author{C. J. Halcrow$^*$}

\author{C. King$^\dagger$}

\author{N. S. Manton$^\ddagger$}

\affiliation{Department of Applied Mathematics and Theoretical Physics, University of Cambridge, Wilberforce Road, Cambridge CB3 0WA, United Kingdom}

\date{\today}

\begin{abstract}
We calculate the low-lying spectrum of the $^{16}$O nucleus using an $\alpha$-cluster model which includes the important tetrahedral and square configurations. Our approach is motivated by the dynamics of $\alpha$-particle scattering in the Skyrme model. We are able to replicate the large energy splitting that is observed between states of identical spin but opposite parities, as well as introduce states that were previously not found in other cluster models, such as a $0^-$ state. We also provide a novel interpretation of the first excited state of $^{16}$O and make predictions for the energies of $6^-$ states that have yet to be observed experimentally.
\end{abstract}


\maketitle

\emph{Introduction} -- The energy spectrum of $^{16}$O has posed a challenge to nuclear physicists for decades. Shell model calculations \cite{SkPerr}, ab initio calculations \cite{abin} and cluster models \cite{TetVib} agree that the $0^+$ ground state arises from a tetrahedral arrangement of $\alpha$-particles. However, the first excited state of $^{16}$O, which also has spin-parity $0^+$, has been the subject of much debate. Ab initio calculations suggest this state is strongly linked to a square configuration of $\alpha$-particles, whereas an analysis of the vibrational modes of the tetrahedron implies that the state is of a breathing nature. The vibrational approach of \cite{TetVib} studies a local harmonic approximation around the tetrahedron, leading to a degeneracy between some states with equal spin but opposite parity, which is not realised in the experimental energy spectrum. In this Letter we provide a global analysis of a vibrational space that includes the tetrahedral and square configurations. This global approach lifts the aforementioned degeneracy and provides a new interpretation of the first excited state as a quantum superposition of these different configurations. It also allows us to construct a spin $0$ state with negative parity -- a first for $\alpha$-cluster models.

\begin{figure}[bp!]
	\includegraphics[width=17.8cm]{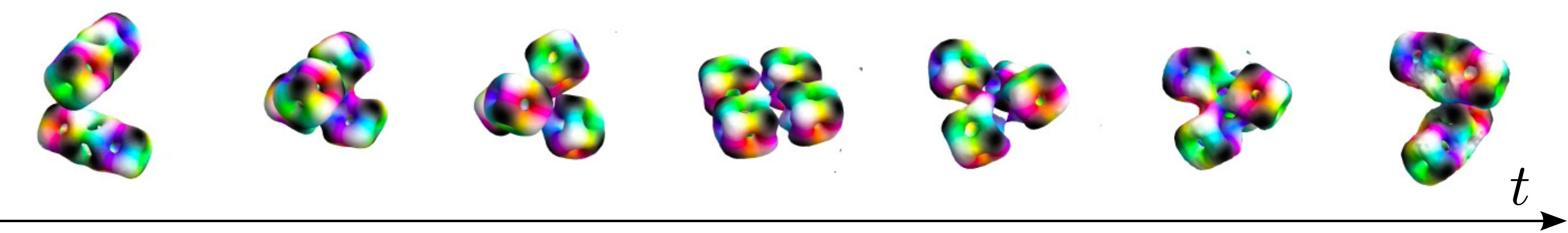}
	\caption{A scattering mode of four $\alpha$-particles in the Skyrme model. Each time step shows a surface of constant energy density which is coloured according to the field value as in \cite{108}.}
	\label{fig:Scattering}
\end{figure}

The Skyrme model \cite{Sk} reproduces $\alpha$-particle clustering \cite{B4build,Hoyle} and contains the tetrahedral and square configurations seen in conventional cluster models of $^{16}$O. The key advantage of the Skyrme model is that it provides dynamics for the clusters, which we use to construct our vibrational space. There is a dynamical mode, shown in Fig. \ref{fig:Scattering}, connecting the tetrahedral and square configurations. Two pairs of $\alpha$-particles approach each other and form a tetrahedron, which flattens out into a square, before reopening into the dual tetrahedron and then breaking into two pairs of $\alpha$-particles again. There are three of these modes passing through each tetrahedron, corresponding to the three pairs of opposite edges. Note that these modes are an extension of the $E$ vibration studied in \cite{TetVib}. If one starts at the tetrahedron and excites each of these modes equally then they will cancel out. Therefore  these three modes only generate a two-dimensional space of configurations which forms our vibrational space. 

\begin{figure}[t]
	\includegraphics[height=4cm]{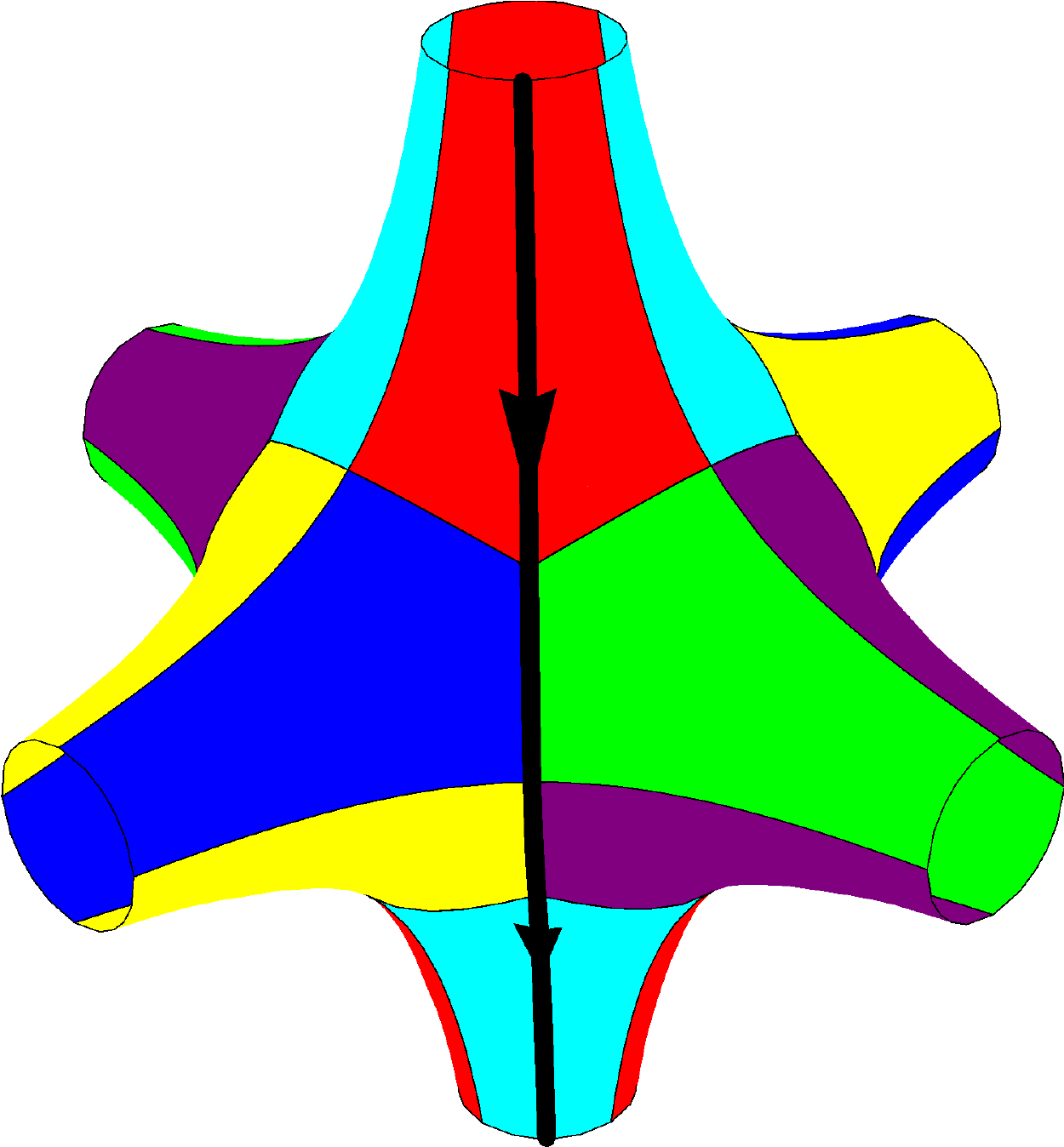}
	\caption{The $\alpha$-particles are restricted to lie on a surface with six punctures. Regions with the same colouring are related by $D_2$ symmetry. The scattering mode in Fig. \ref{fig:Scattering} is represented by the thick black line.}
	\label{fig:6pun}
\end{figure}

The degrees of freedom in this vibrational space are the positions of the $\alpha$-particles, which lie on a surface. To account for the asymptotics seen in Fig. \ref{fig:Scattering}, this surface must stretch out to infinity in six directions as in Fig. \ref{fig:6pun}. Each configuration has $D_2$ symmetry and hence if one $\alpha$-particle is at $\boldsymbol{x}=(x,y,z)$, the others are at $(x,-y,-z),(-x,y,-z)$ and $(-x,-y,z)$. This means we may focus on one quarter of the surface which we denote by $\mathcal{M}$. 

Having constructed the vibrational space we can now quantise the system using the scheme laid out in \cite{AttDeu,B7vib}. The total configuration space is $\mathcal{M} \times SO(3)$, which allows for rotations of each configuration. The quantum Hamiltonian is
\begin{equation} \label{Schro}
\hat{H} = -\frac{\hbar^2}{2}\Delta + V(\boldsymbol{x}) \, ,
\end{equation}
where $V(\boldsymbol{x})$ is the static energy of the configuration on

\onecolumngrid

\pagebreak
\twocolumngrid

\noindent $\mathcal{M}$ with an $\alpha$-particle at $\boldsymbol{x}$, and the kinetic operator is proportional to the Laplace--Beltrami operator
\begin{equation}
\Delta = \det(g)^{-\frac{1}{2}}\partial_i\left(\det(g)^{\frac{1}{2}} g^{ij} \partial_j\right) \, ,
\end{equation}
where $g$ is the metric on $\mathcal{M}\times SO(3)$. The metric is block diagonal and hence the problem splits into vibrational and rotational parts. The total wavefunction is separable and can be written as
\begin{equation}
\ket{\Psi}  = \sum_{L_3} \phi_{L_3}(\boldsymbol{x}) \ket{J L_3} \, ,
\end{equation}
where $\phi$ is the vibrational wavefunction and $\ket{J L_3}$ are the rigid-body angular momentum states with spin $J$ and body-fixed angular momentum projection $L_3$. In addition, the linear combination of states occurring in $\ket{\Psi}$ must be $D_2$ invariant.

\emph{The vibrational problem} -- The rescaled Schr\"odinger equation for the vibrational wavefunction is
\begin{equation} \label{VibSchro}
-\Delta_{\rm vib}\,\phi + V(\boldsymbol{x})\phi = (E-E_J) \phi \, ,
\end{equation}
where $E$ is the total energy of $\ket{\Psi}$ and $E_J$ is its rotational energy. $E_J$ involves the moments of inertia of the configurations, which depend on the vibrational coordinates $\boldsymbol{x}$. However, for now, we consider them to be constant to simplify the calculation of the vibrational energy.

\begin{figure}[h]
$\begin{array}{l}
 \includegraphics[scale=0.35,keepaspectratio=true]{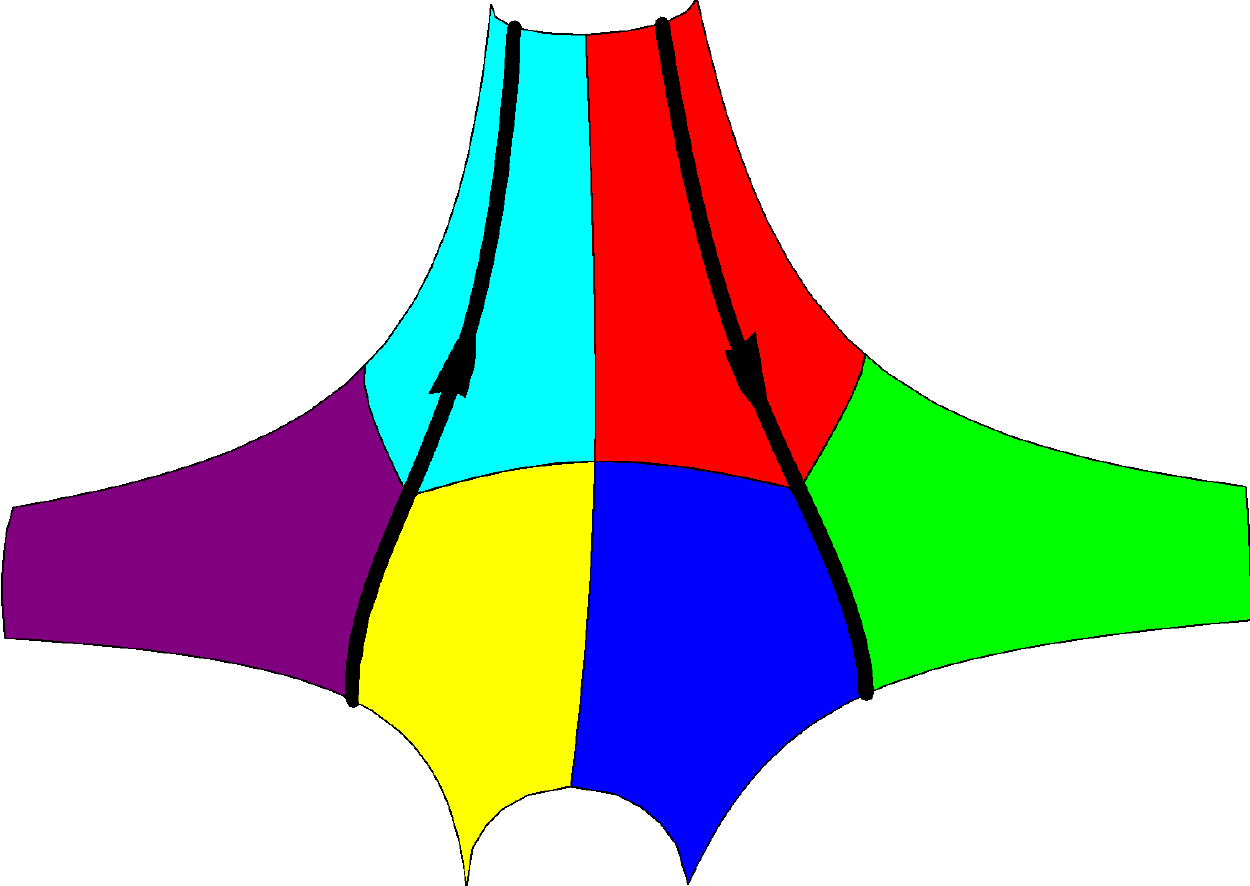}
 \label{fig:equiv}
\end{array}$
$\cong$
$\begin{array}{l}
 \includegraphics[scale=0.25,keepaspectratio=true]{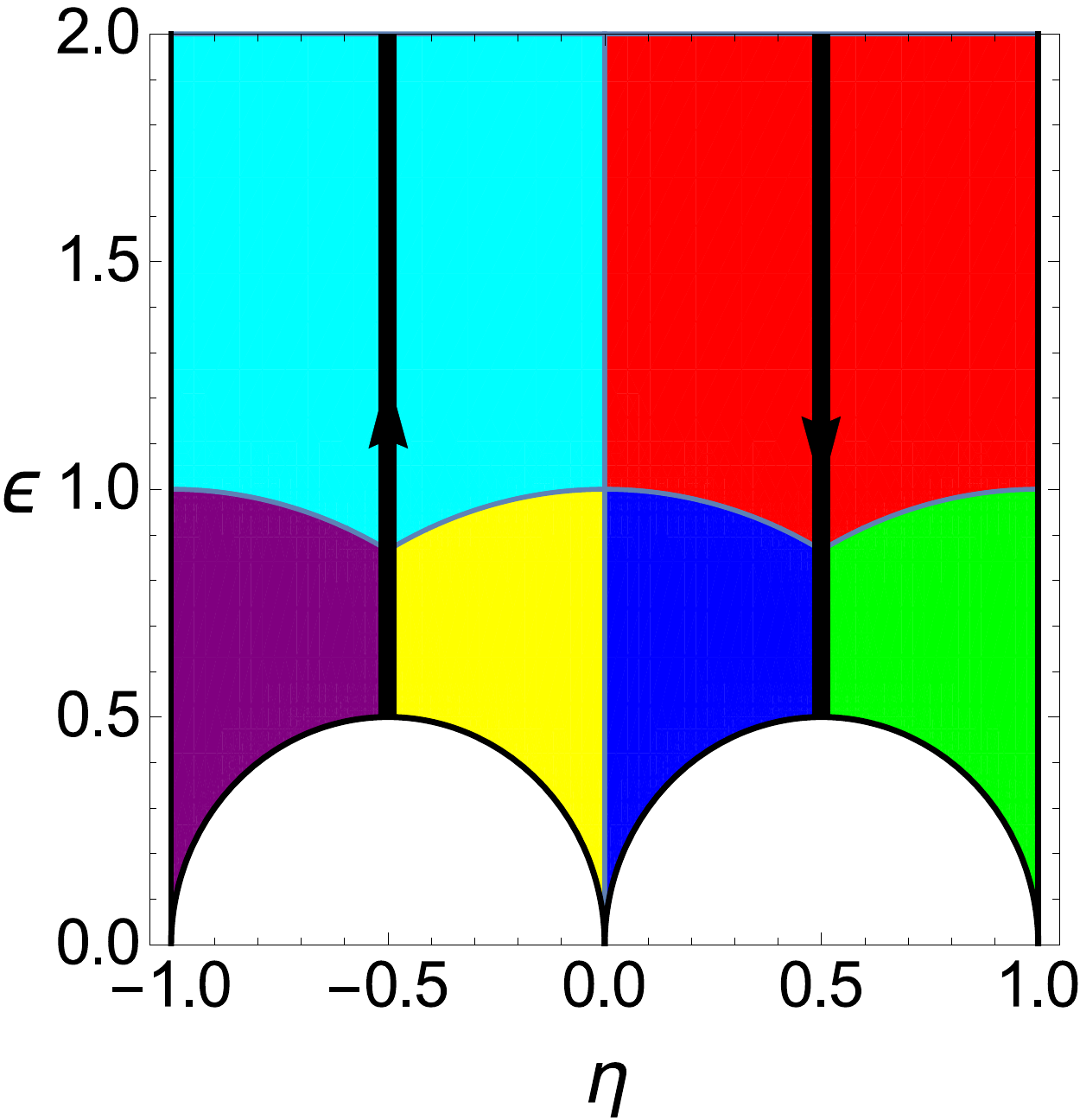}
\end{array}$
\caption{The relation between $\mathcal{M}$ (left) and $\mathcal{F}$ (right). Tetrahedral configurations are at the points where three coloured regions meet while the square configurations are at points where four coloured regions meet. The scattering mode in Fig. \ref{fig:Scattering} is represented by the thick black lines.}
\label{equiv}
\end{figure}

To solve \eqref{VibSchro} we must first model the metric on our space $\mathcal{M}$. We will use the $6$-punctured sphere with constant negative curvature to approximate $\mathcal{M}$. The metric is simple once we map $\mathcal{M}$ onto a sub-domain $\mathcal{F}$ of the complex upper half plane $\mathbb{H}$, which is conformally equivalent. The sub-domain for this problem is $\mathcal{F} \equiv \mathbb{H}/\Gamma(2)$, where $\Gamma(2)$ is a modular subgroup, and the map from $\mathcal{F}$ to $\mathcal{M}$ is known \cite{UniBook}. The conformal equivalence between $\mathcal{M}$ and $\mathcal{F}$ is shown in Fig. \ref{equiv}. Defining $\zeta = \eta + i \epsilon$ as the complex coordinate on $\mathbb{H}$ with $\epsilon$ positive, the metric is proportional to $\epsilon^{-2}\left(d\eta^2+d\epsilon^2\right)$ giving rise to the kinetic operator
\begin{equation}
-\Delta_{\rm vib} = -\epsilon^2\left(\frac{\partial^2}{\partial \eta^2}+\frac{\partial^2}{\partial \epsilon^2}\right) \, .
\end{equation}

The $6$-punctured sphere has cubic symmetry $O$ and hence $\mathcal{M}$ has $O/D_2 \cong S_3$ symmetry, where $S_3$ is the permutation group of three objects. $S_3$ can act on $\mathcal{M}$ or $\mathcal{F}$, permuting the coloured regions seen in Fig. \ref{equiv}. In addition, parity acts on $\mathcal{M}$ as
\begin{equation}\label{par}
\boldsymbol{x} = (x,y,z) \to (-x,-y,-z) \equiv (x,-y,z)
\end{equation}
where we have used the $D_2$ symmetry in the equivalence. This corresponds to $\eta \to -\eta$ on $\mathcal{F}$. Hence the vibrational wavefunctions fall into representations of $S_3$ and parity. 

The group $S_3$ has three irreducible representations which we shall now review \cite{reps}. The trivial representation maps all elements to $+1$, whereas the sign representation maps the identity and $3$-cycles to $+1$, and transpositions to $-1$. $S_3$ also admits the natural representation, which acts on triplets $(u,v,w)$ via permutation. However, this is not irreducible because the combination $u+v+w$ is invariant under any element of $S_3$. A $2$-dimensional irrep, called the standard representation, is obtained by letting $S_3$ act on triplets $(u,v,w)$ such that $u+v+w=0$. We can now apply elements of $S_3$ in two ways; either directly on the coordinates of $\mathcal{F}$ or via a representation on the solutions $\phi$. These two methods have to agree, which enables us to the find boundary conditions that each type of wavefunction must obey.

Our choice of potential $V$ is motivated by cluster models which find that the tetrahedral configuration has the lowest energy \cite{alpha}. Going towards the square or asymptotic configurations leads to a rise in potential energy. In addition, we would like a potential for which \eqref{VibSchro} is soluble. A natural candidate is
\begin{equation} \label{pot}
V(\eta,\epsilon)=\epsilon^2\left(\omega^2\left(\eta-\frac{1}{2}\right)^2+\mu^2\right) \, ,
\end{equation}
where $\omega$ and $\mu$ are constant parameters and the $\epsilon^2$ factor means that solutions of \eqref{VibSchro} are separable in $\eta$ and $\epsilon$. This formula only applies in the top right region of $\mathcal{F}$ and the potential elsewhere can be found by defining $V$ to take the same value at points related by $S_3$.

The solutions of equation \eqref{VibSchro} are of the form
\begin{equation}
\phi(\eta,\epsilon) = \sum_n a_n H_n(\eta)G_n(\epsilon)\sqrt{\epsilon} \, ,
\end{equation}
where $H_n$ satisfies the harmonic oscillator problem on an interval and $G_n$ are modified Bessel functions. The coefficients $a_n$ are determined by matching conditions on the boundaries between regions of $\mathcal{F}$.

\emph{Rovibrational states} -- The vibrational wavefunctions must be combined with spin states in order to form rovibrational states. The combinations that are permitted depend on the representation that the vibrational wavefunction falls into. So far $S_3$ has only acted on vibrational space, but each element is also equivalent to a rotation. For example, the transposition $(1 \, 2)$ is equivalent to a $\frac{\pi}{2}$ rotation about the $(0,0,1)$ axis, whereas the $3$-cycle $(1 \, 2 \, 3)$ is equivalent to a $\frac{2\pi}{3}$ rotation about the $(1,1,1)$ axis. The corresponding rotation operators are given by exponentiating suitable combinations of the body-fixed angular momentum operators $\hat{L}_i$, which act on the spin states. These states, just like the vibrational wavefunctions, fall into representations of $S_3$. To form a composite rovibrational state, the individual vibrational and rotational states must fall into the same representation.

\begin{figure}[htb!]
\centering
\includegraphics[scale=1]{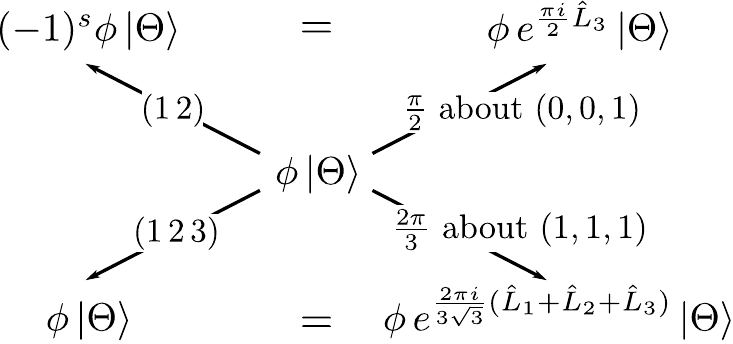}
\caption{The action of elements of $S_3$ on vibrational (left) and spin (right) states in the trivial and sign representations. The parameter $s$ is equal to $0$ and $1$ for the trivial and sign representations respectively.}
\label{S3sing}
\end{figure}

Rovibrational states in the trivial or sign representation are separable, and the alternative ways of applying $S_3$ provide the constraints shown in Fig. \ref{S3sing}. The spin states satisfying these are exactly those which are compatible with a tetrahedrally symmetric configuration of $\alpha$-particles. The trivial representation allows for spin $J=0,4,6,\ldots$ while the sign representation allows for spin $J=3,6,\ldots$. Rovibrational states of either parity are permitted for all of these spins, but negative (positive) parity solutions in the trivial (sign) representation have rather high energy since the vibrational wavefunctions must vanish at the tetrahedral configuration.

\begin{figure}[htb!]
\centering
\includegraphics[scale=0.217]{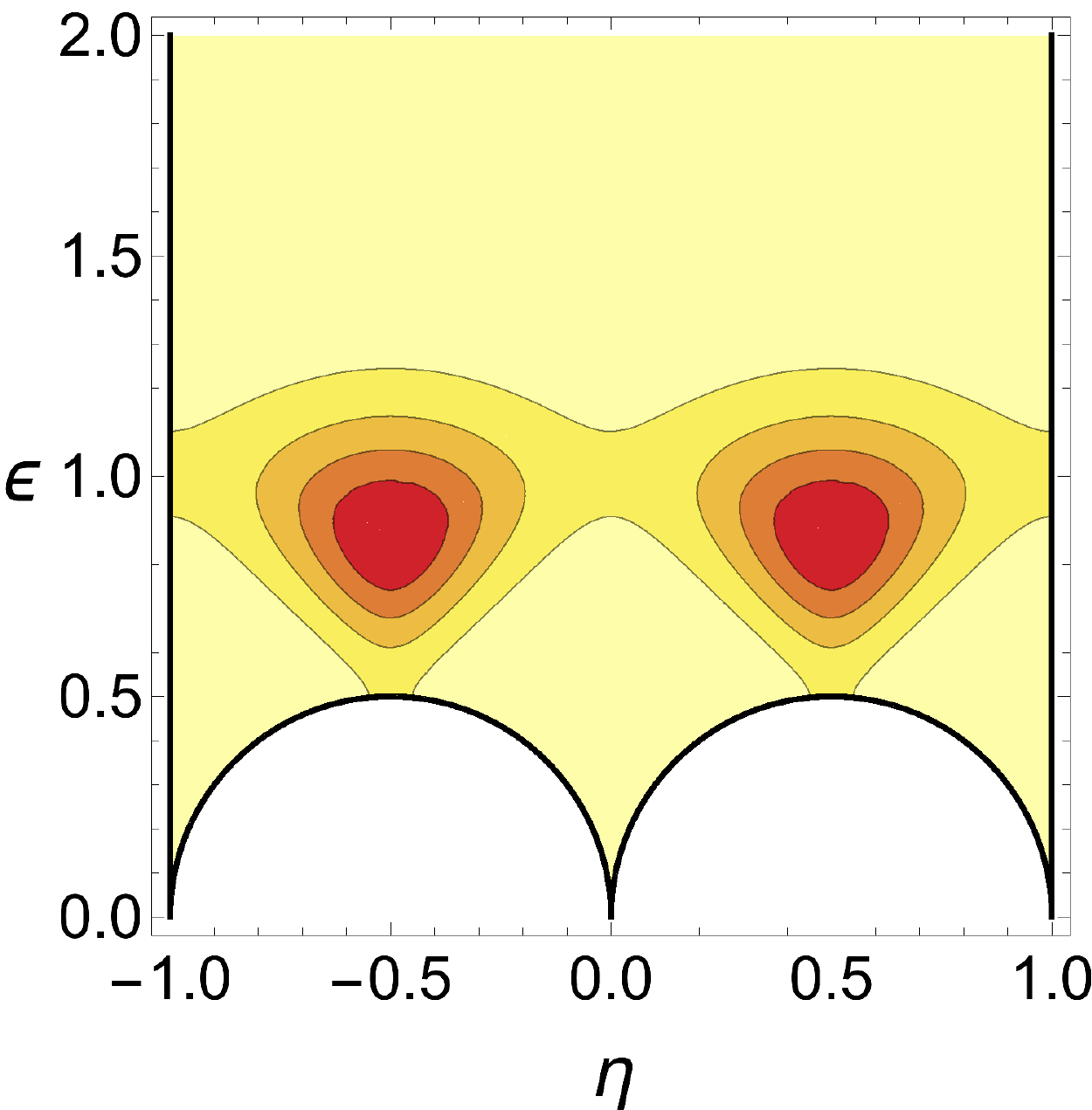}
\includegraphics[scale=0.217]{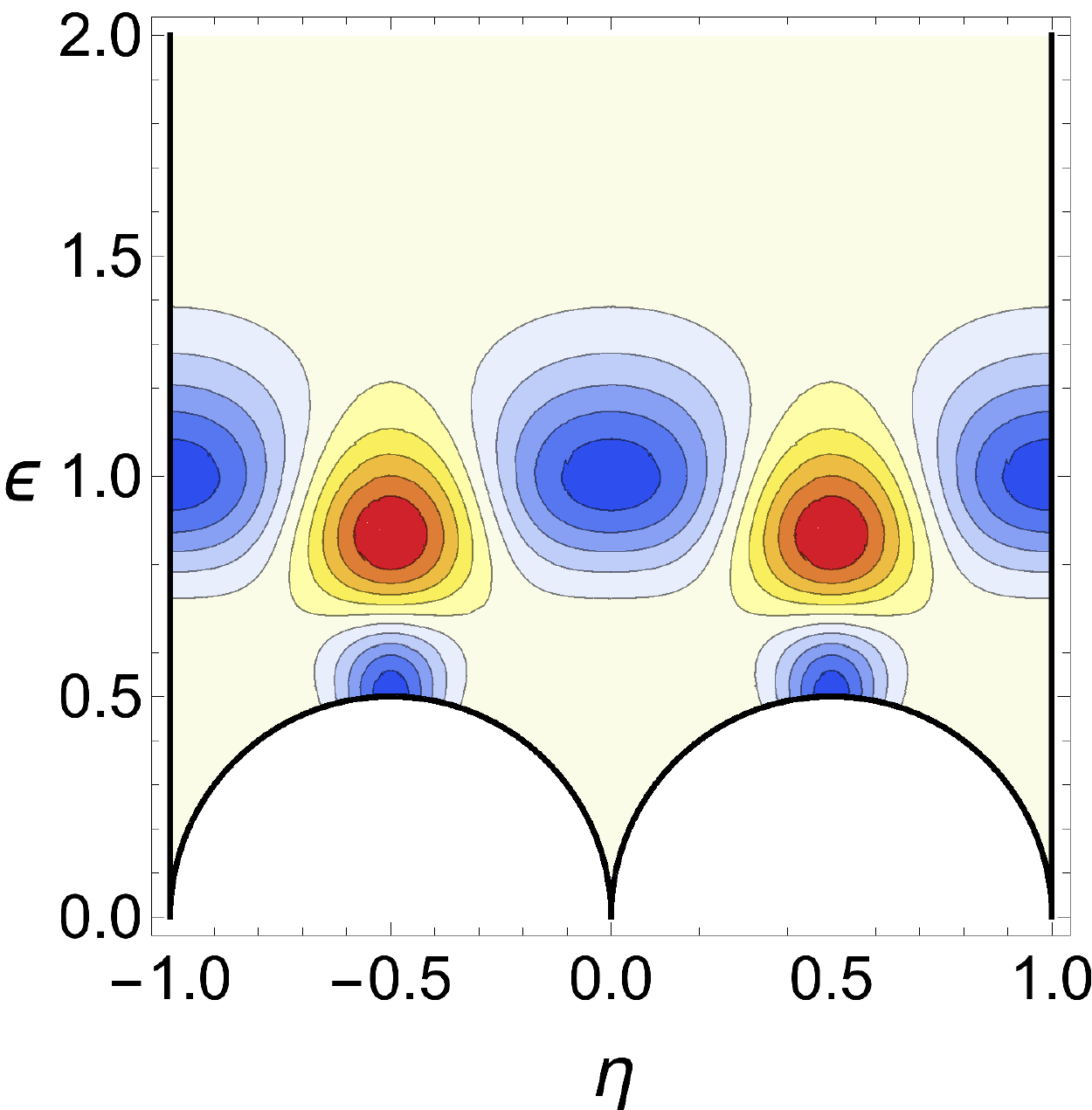}
\includegraphics[scale=0.217]{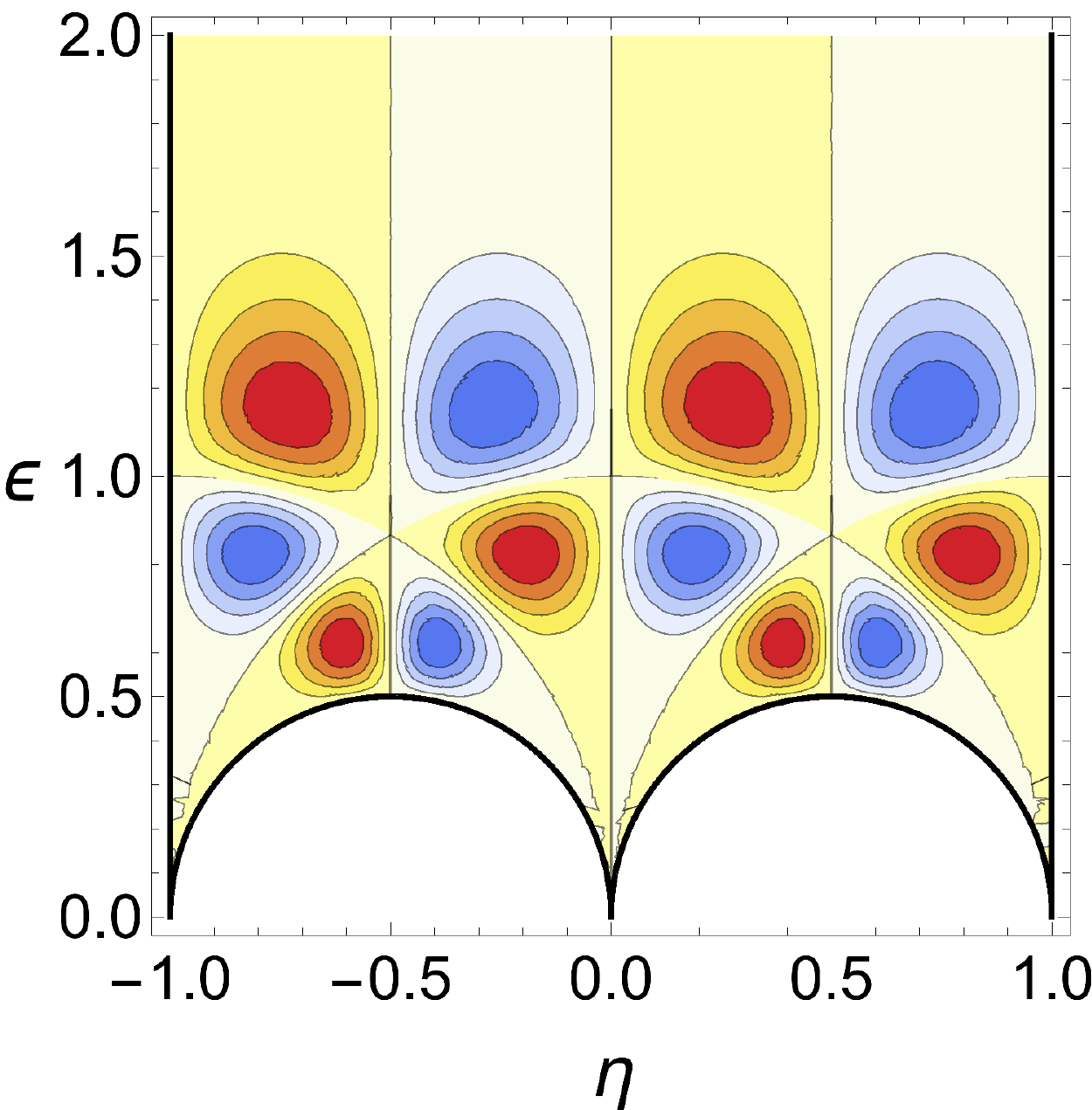}
\caption{Vibrational wavefunctions which lie in the trivial representation. From left to right: the ground state, the first excited state, and the lowest-lying state with negative parity.}
\label{trivrep}
\end{figure}

The two lowest-energy vibrational wavefunctions in the trivial representation are displayed in Fig. \ref{trivrep} (left and middle), where red and blue colours correspond to positive and negative values respectively. We identify these solutions, when combined with the $\ket{0,0}$ spin state, with the two lowest $0^+$ states in the experimental spectrum of $^{16}$O.  The ground state is loosely focused around the two tetrahedral configurations in agreement with other models. The excited state has probability maxima at the three square configurations, and at the two tetrahedral configurations, of roughly equal magnitude. Hence we deduce that a global analysis, including both tetrahedral and square configurations, appears essential to explain the structure of the excited $0^+$ state.

Each configuration is related to another, its dual configuration, by parity. If these configurations have a reflection symmetry then they are also related by a rotation. If the spin state is $\ket{0,0}$, which is invariant under any rotation, the corresponding vibrational wavefunction must take the same value at the configuration and its dual.  This is automatic for positive parity vibrational wavefunctions. Negative parity wavefunctions are also permitted if they vanish at all configurations with a reflection symmetry, including the tetrahedral and square configurations. This vanishing condition is already a natural consequence of the combination of negative parity with the trivial representation. Therefore negative parity vibrational wavefunctions, such as the right-most wavefunction in Fig. \ref{trivrep}, can be combined with the $\ket{0,0}$ spin state to give a $0^-$ state. This is the first time such $0^-$ states have been accommodated in either $\alpha$-cluster models or the Skyrme model.

\begin{figure}[htb!]
\centering
\includegraphics[scale=0.217]{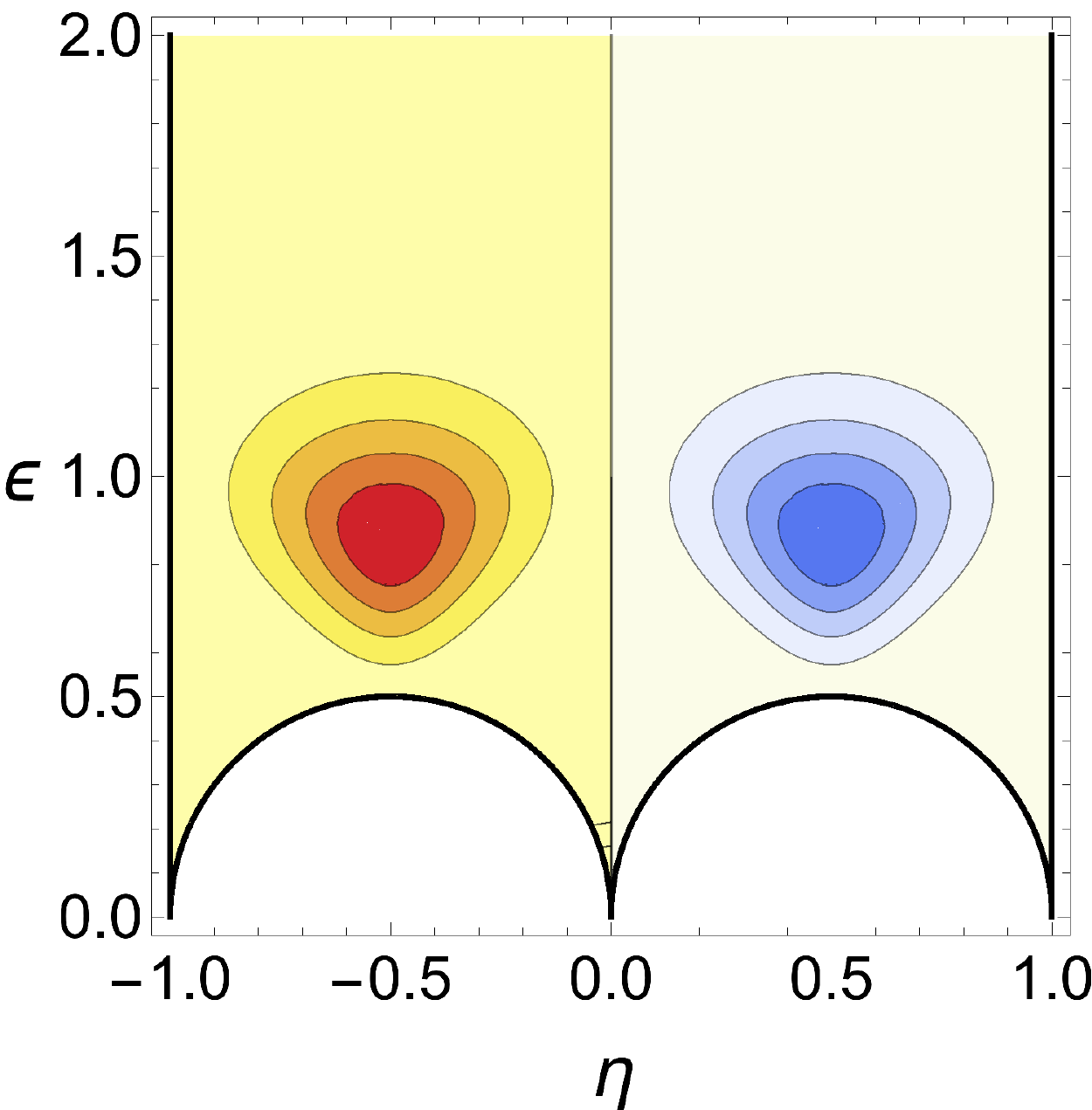}
\includegraphics[scale=0.217]{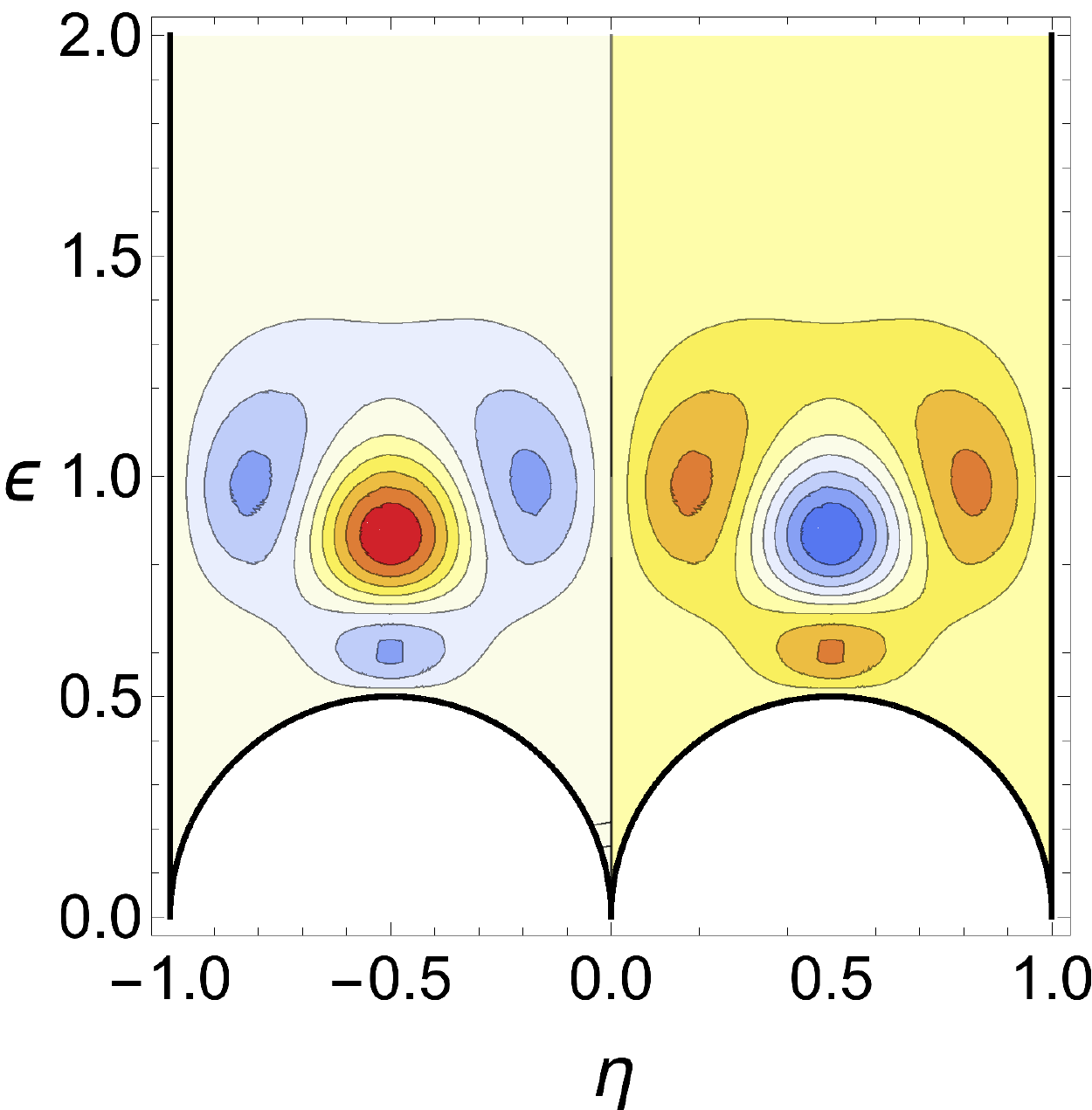}
\includegraphics[scale=0.217]{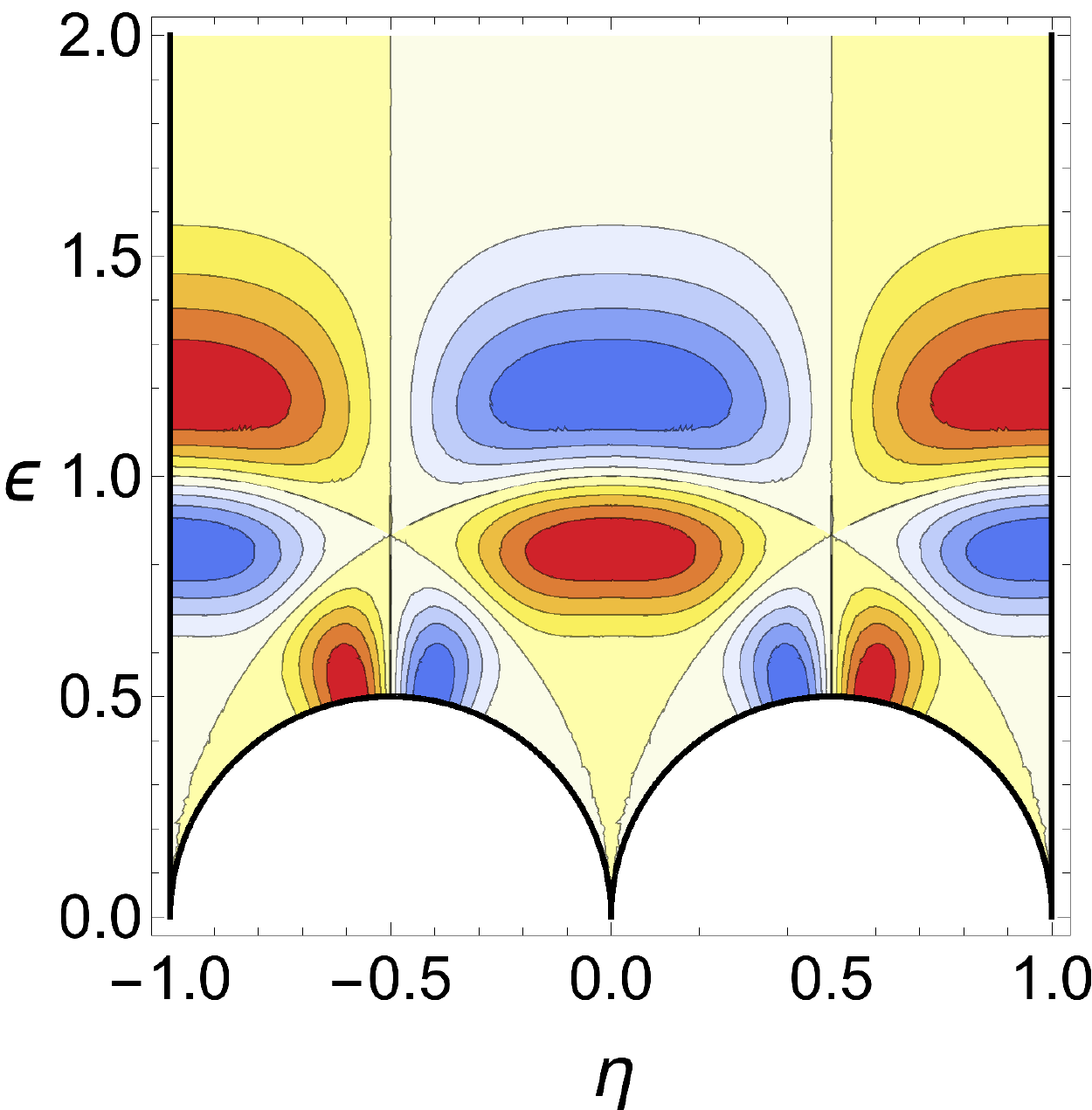}
\caption{Vibrational wavefunctions which lie in the sign representation. From left to right: the lowest-energy state, the first excited state, and the lowest-lying state with positive parity.}
\label{signrep}
\end{figure}

Vibrational wavefunctions in the sign representation with negative (positive) parity are not too different from those in the trivial representation with positive (negative) parity. The sign representation wavefunctions are displayed in Fig. \ref{signrep} and the similarities with those in Fig. \ref{trivrep} are manifest. The two wavefunctions on the left give rise to $3^-$ states while the right-most wavefunction gives rise to a $3^+$ state of rather high energy.

\begin{figure}[htb!]
\centering
\includegraphics[scale=1]{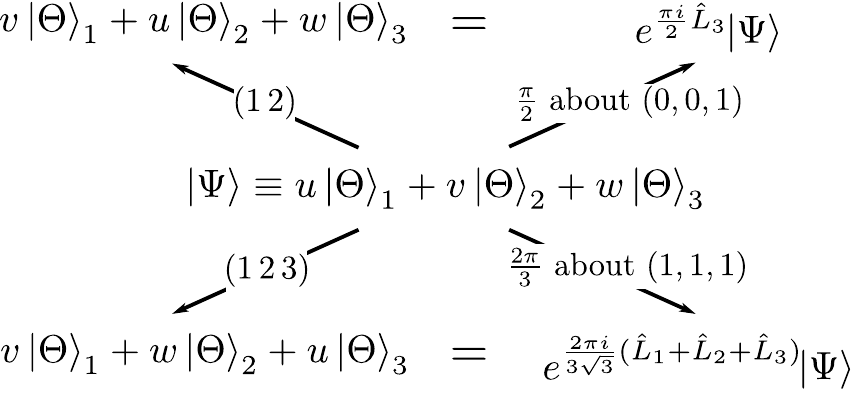}
\caption{The action of elements of $S_3$ vibrational and rotational states in the standard representation.}
\label{S3doub}
\end{figure}

Rovibrational states in the standard $2$-dimensional representation are most easily presented using a $3$-dimensional set of spin states $\ket{\Theta}_i$ which $S_3$ permutes as shown in Fig. \ref{S3doub}.

\begin{figure}[htb!]
\centering
\includegraphics[scale=0.162]{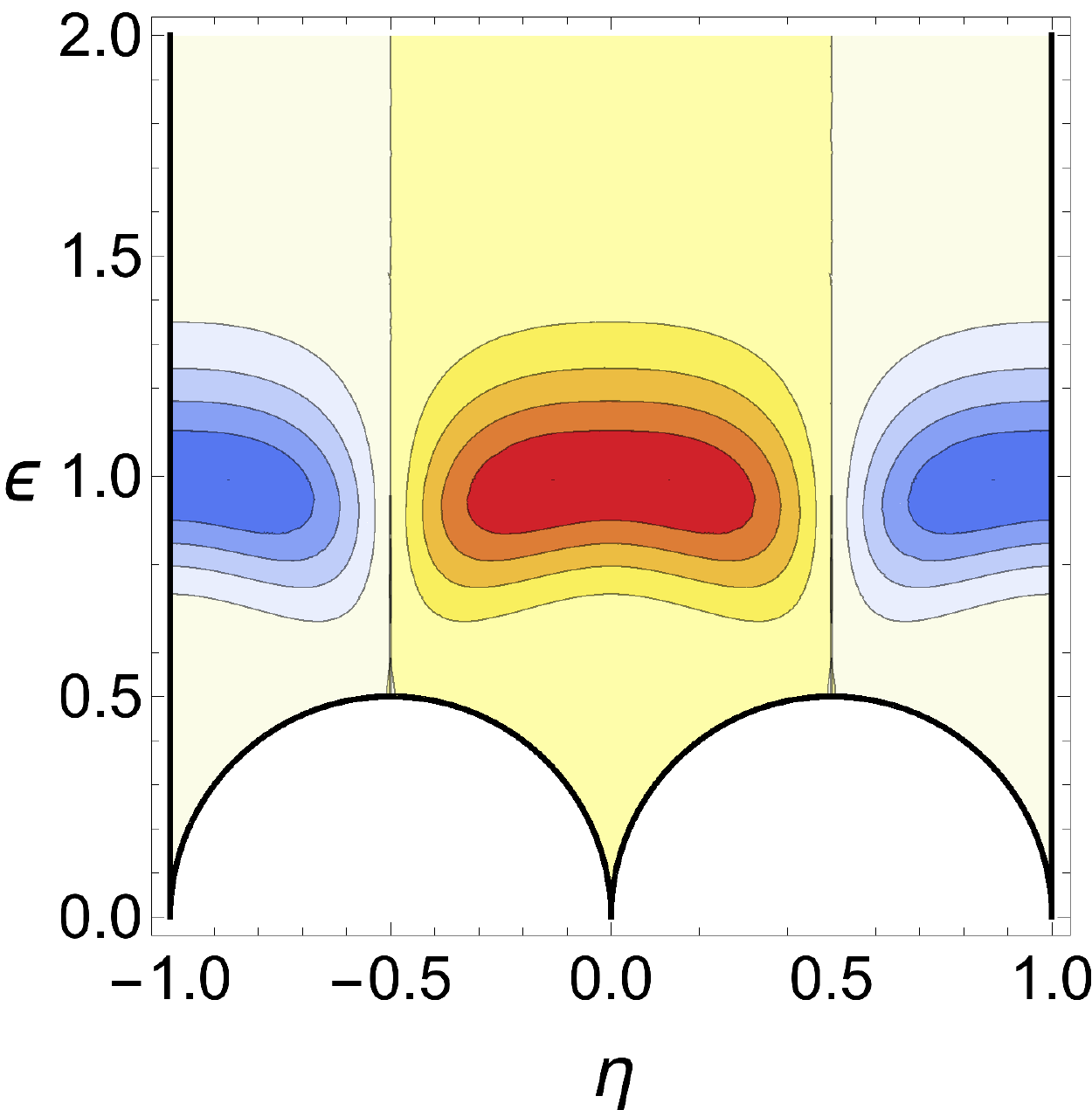}
\includegraphics[scale=0.162]{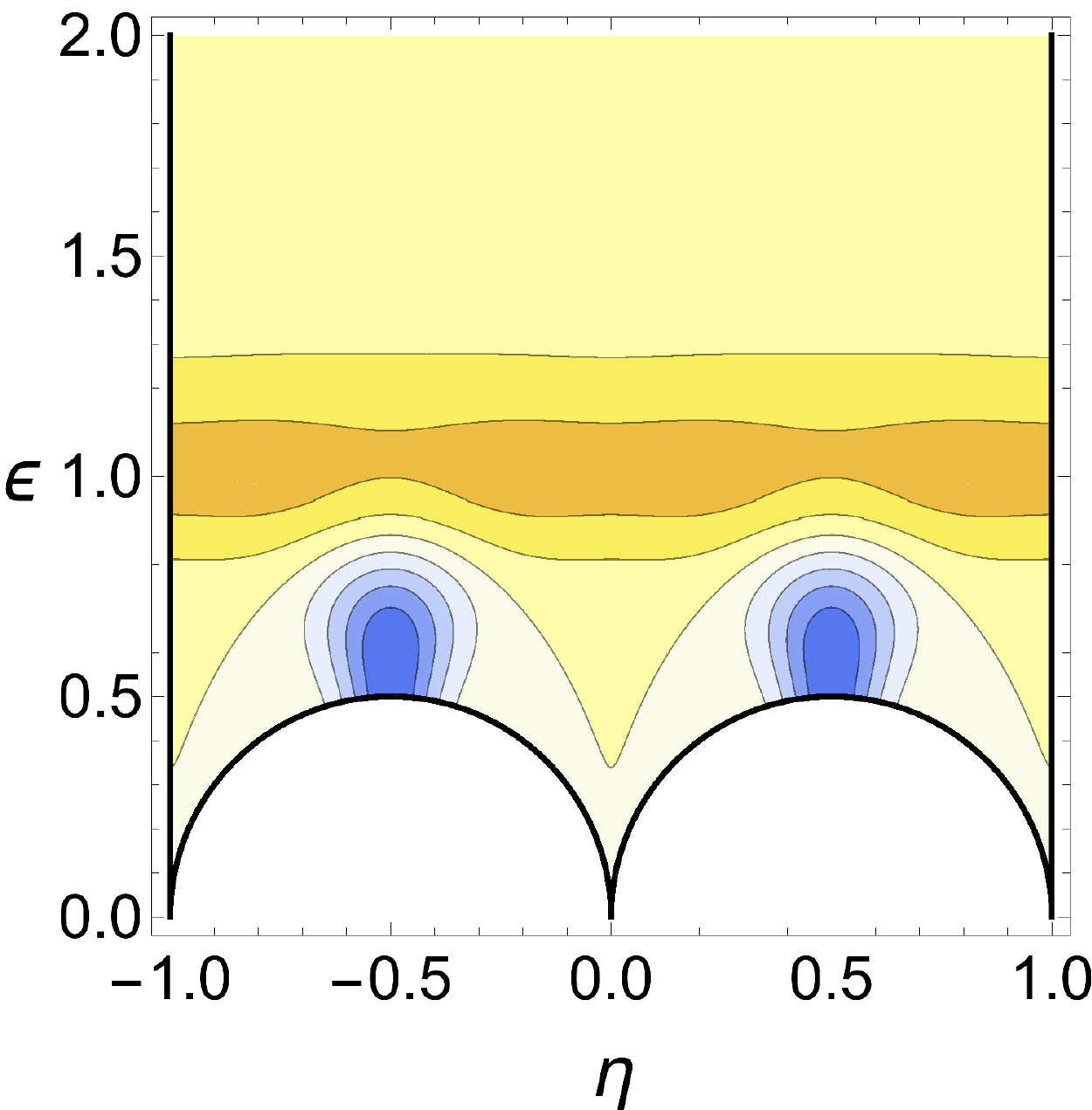}
\includegraphics[scale=0.162]{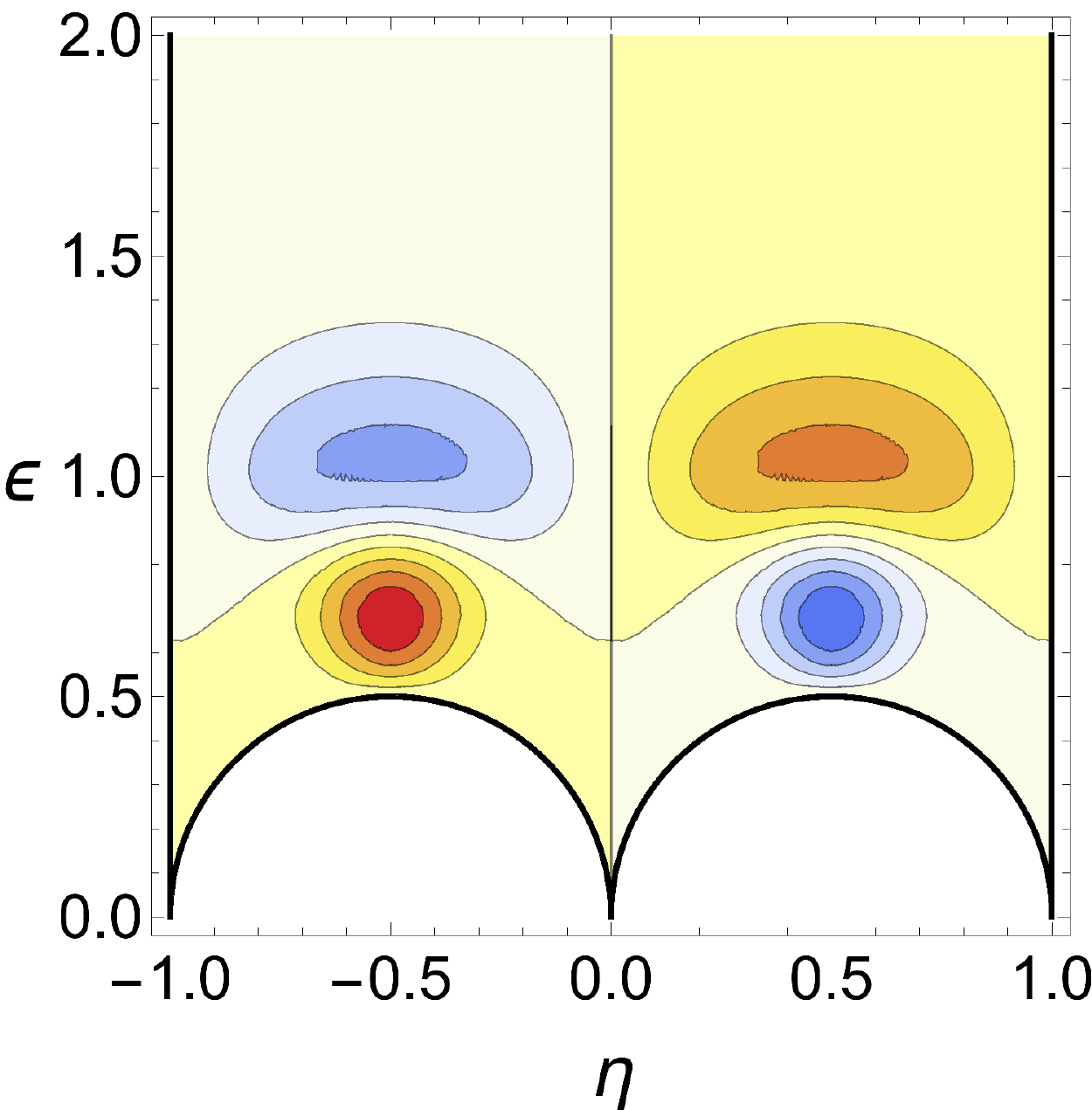}
\includegraphics[scale=0.162]{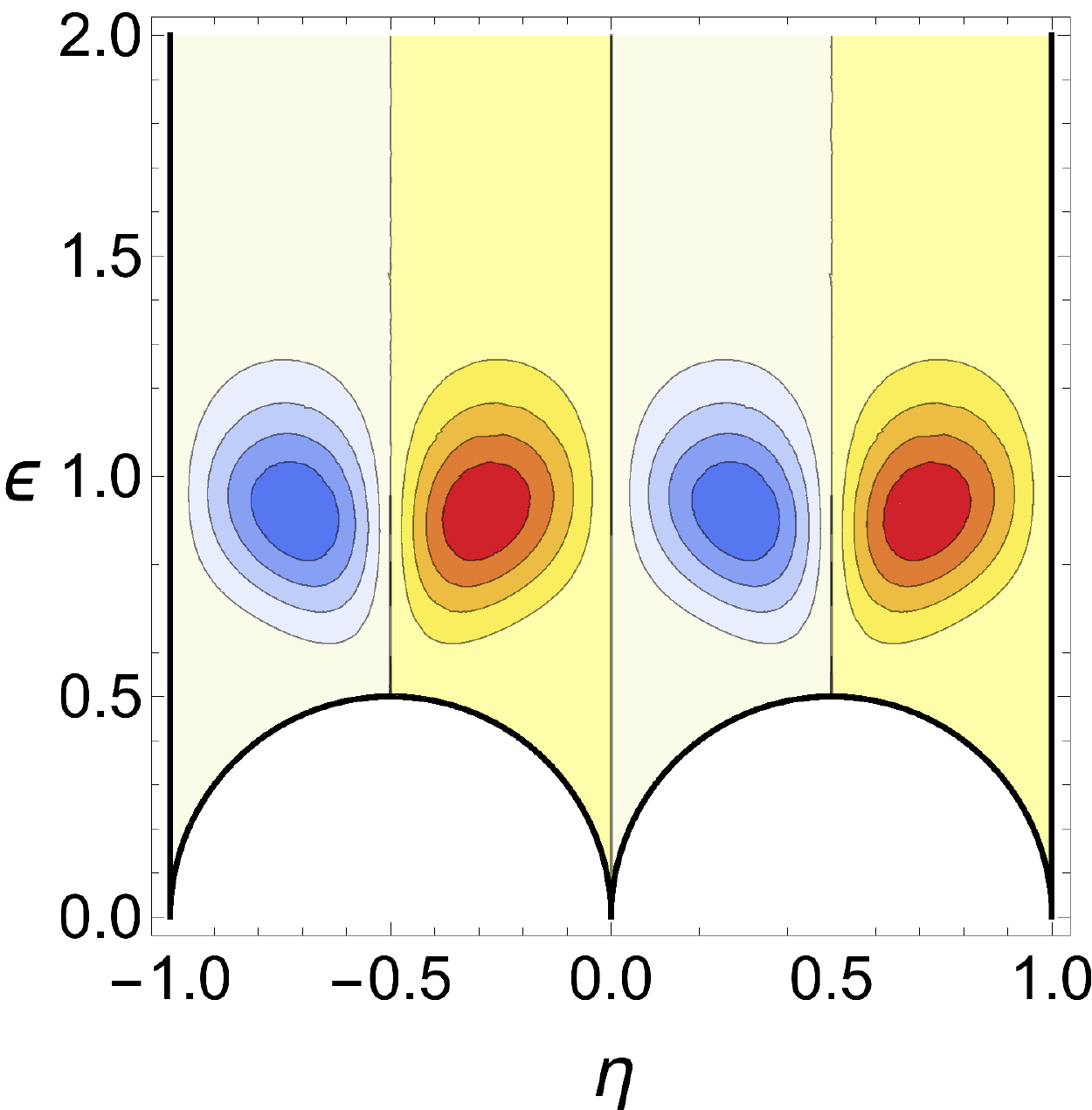}
\caption{The lowest-energy vibrational wavefunctions which lie in the standard representation. From left to right: the function $u-v$ with positive parity, the function $u+v$ with positive parity, the function $u-v$ with negative parity and the function $u+v$ with negative parity. }
\label{standpos}
\end{figure}

For spin $2$, $\ket{\Theta}_i$ is the state with zero projection on the $i^{\text{th}}$ axis, and these states satisfy
\begin{equation} 
\hat{L_i}\ket{\Theta}_i=0 \, \text{ and } \, \ket{\Theta}_1+\ket{\Theta}_2+\ket{\Theta}_3=0 \, .
\end{equation}
One may check that they transform into each other under rotations so that the equalities shown in Fig. \ref{S3doub} are satisfied. The total wavefunction is then
\begin{align}
 \ket{\Psi} &= u \ket{\Theta}_1 + v \ket{\Theta}_2 + w \ket{\Theta}_3 \nonumber  \\
&=\frac{\sqrt{3}}{2 \sqrt{2}}(u-v)\left(\ket{2,2}+\ket{2,-2}\right)-\frac{3}{2}(u+v)\ket{2,0},
\end{align}
rewritten in terms of the usual spin states, with definite $L_3$ eigenvalues. Notice how the rovibrational state can be expressed in terms of a $2$-dimensional basis of spin states, as would be expected in the standard representation. We can construct similar states for spin-parity $J^P = 2^{\pm},4^{\pm},5^{\pm},6^{\pm},\ldots$. The lowest-energy positive and negative parity vibrational wavefunctions are displayed in Fig. \ref{standpos}. The negative parity states have higher energy than the positive parity states since they are more constrained, having to vanish at all square configurations. As with the other representations, there are further vibrationally excited states which we have calculated though not displayed.

\emph{The energy spectrum} -- To find the vibrational wavefunctions in \eqref{VibSchro} we neglected the fact that the moments of inertia depend on the vibrational coordinates. We now include the energy contribution from this dependence by calculating $\bra{\Psi}\hat{H}\ket{\Psi}$ using an approximate inertia tensor that interpolates between the known values for the Skyrme model configurations shown in Fig. \ref{fig:Scattering} \cite{EMtrans}. This is equivalent to using first order perturbation theory, which is justified as the energy gaps between states in the same representation are large.
\begin{figure}[htb!]
\centering
\includegraphics[width=8.6cm]{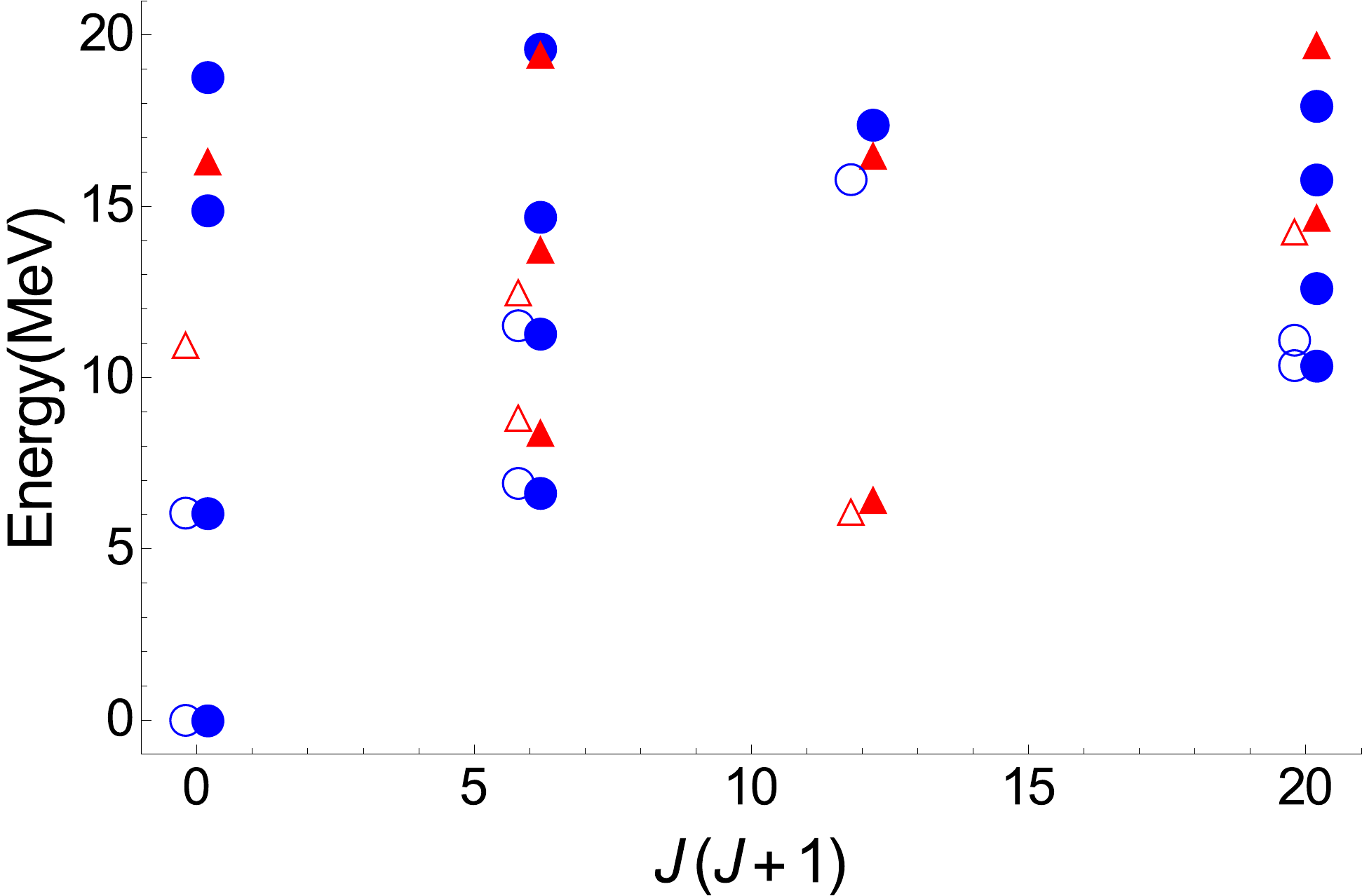}
\caption{The energy spectrum of our model. Calculated states of positive parity (solid circles) and negative parity (solid triangles) are plotted. Where the identification with an experimentally observed state \cite{O16Exp} is clear, we also plot these (hollow symbols).}
\label{spec}
\end{figure}

The ground state is fixed at $0$ MeV, with the excited $0^+$ and lowest $4^+$ state being used to scale the vibrational and rotational energy units respectively. The two remaining parameters ($\omega$ and $\mu$) are chosen to give a good fit for the rest of the states. The calculated spectrum up to $20$ MeV is shown in Fig. \ref{spec} and matches the experimental spectrum well.

The lowest-lying $2^+$ and $2^-$ states have the correct ordering, with a predicted energy gap of $1.8$ MeV which is close to the experimentally observed gap of $1.96$ MeV. This gap is caused by the vibrational wavefunctions having significantly different energies, due to their opposite parities. A global analysis is essential to describe this gap. The lowest $0^+$, $3^-$ and $4^+$ states still form a rotational band, despite the fact that the $3^-$ state has a different vibrational wavefunction (compare Fig. \ref{signrep} with Fig. \ref{trivrep} (left)). For our choice of parameters in the potential, these vibrational wavefunctions have similar energies.

The energy of the $0^-$ state in Fig. \ref{trivrep} is $16.35$ MeV which is significantly larger than the lowest experimentally observed $0^-$ state which has energy $11.0$ MeV. In our calculation we have used the potential \eqref{pot} which diverges asymptotically; however, the configuration energy should flatten out as we approach the two separated pairs of $\alpha$-particles ($\epsilon \to \infty$). Taking this into account would reduce the vibrational energy of all states but have a larger effect on highly excited states such as the $0^-$.

We find a $6^+$ state in the trivial $S_3$ representation at $21.7$ MeV which agrees with an experimentally observed state at $21.6$ MeV. In addition, we predict two $6^-$ states: one at $22.2$ MeV from the sign representation and one at $27.1$ MeV from the standard representation. Negative parity spin $6$ states have not yet been observed. 

Our model cannot describe several low-lying states as we only include the $D_2$ symmetric configurations, corresponding to the $E$ vibration. Inclusion of the $F$ vibration of the tetrahedron (which breaks $D_2$ symmetery) would give rise to the experimentally observed $1^-,2^+,3^\pm,\dots$ states while the breather mode could explain a $0^+, 3^-,4^+,\dots$ band higher up in the spectrum \cite{TetVib}. The general formalism of vibrational quantisation, inspired by the Skyrme model, can be used to include these vibrations and we hope to examine them in the future. We also hope to study the electromagnetic transitions between our states, extending the work in \cite{EMtrans}.

\emph{Conclusion} -- We have considered an $\alpha$-cluster model for $^{16}$O with novel dynamics motivated by the Skyrme model. Our work allows for $\alpha$-particle configurations with tetrahedral and square symmetry within a two-parameter family of configurations, going beyond the rigid body analysis considered previously \cite{alpha} and also the local analysis of the $E$ vibration in \cite{TetVib}. The $0^+$ ground state is focused around the tetrahedral configuration in agreement with other models, but we provide a novel explanation for the excited $0^+$ state as a superposition of the tetrahedral and square configurations. Our model also allows a $0^-$ state which vanishes at the tetrahedral and square configurations, although these constraints give it rather high energy. We also explain the energy gap between the low-lying $2^+$ and $2^-$ states as being mainly due to their considerably different vibrational wavefunctions. 

\emph{Acknowledgements} -- C. J. H. and C. K. are supported by STFC studentships.

\end{document}